\documentclass[11pt]{article}
\usepackage{amsmath,fullpage,graphicx,booktabs,palatino,epsfig,fancyhdr,color,amssymb,relsize,setspace}

\renewcommand{\headrule}{{\hrule width\headwidth height\noheadrule}}

\newcommand{\appsection}[1]{\let\oldthesection\thesection
\renewcommand{\thesection}{Appendix \oldthesection}
\section{#1}\let\thesection\oldthesection}

\begin{document}

\begin{figure*}[h]
\begin{center}
\vspace{-1.0cm}
\includegraphics[height=4.0cm]{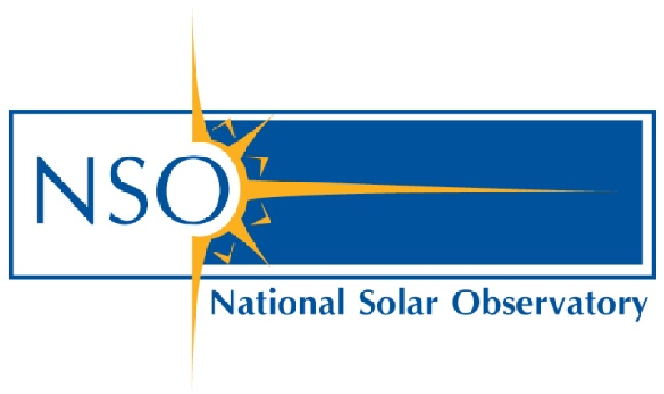}
\end{center}
\end{figure*}

\begin{center}

\vspace{2.0cm}

\begin{LARGE}
\textbf{SOLIS/VSM Polar Magnetic Field Data}
\end{LARGE}

\vspace{3.5cm}

\begin{large}

Luca Bertello \& Andrew R. Marble

\vspace{0.5cm}

National Solar Observatory

\vspace{3.5cm}

\today

\vspace{4.5cm}

\hrule

\vspace{1.0cm}

Technical Report No. \textbf{NSO/NISP-2015-002}

\end{large}

\end{center}

\clearpage
\renewcommand{\headrule}{{\hrule width\headwidth 
height\headrulewidth\vskip0.5cm}}
\fancyhead{}
\fancyhead[L]{SOLIS/VSM Polar Magnetic Field Data}
\fancyhead[R]{\thepage}
\fancyfoot{}
\thispagestyle{fancy}

\vspace*{-0.40cm}

\begin{center}
\begin{Large}
\bf{Abstract}
\end{Large}
\end{center}

\begin{quote}
The Vector Spectromagnetograph (VSM) instrument on the Synoptic Optical Long-term Investigations of the Sun (SOLIS)
telescope is designed to obtain high-quality magnetic field observations in 
both the photosphere and chromosphere by measuring the Zeeman-induced polarization of spectral lines.
With 1$^{\prime \prime}$ spatial resolution (1.14$^{\prime \prime}$ before 2010) and 0.05\AA\ spectral resolution, the VSM
provides, among other products,
chromospheric full-disk magnetograms using the CaII 854.2 nm spectral line 
and both photospheric full-disk vector and longitudinal magnetograms using the FeI 630.15 nm line.
Here we describe the procedure used to compute daily weighted averages of the photospheric radial
polar magnetic field at different latitude bands from SOLIS/VSM longitudinal full-disk
observations. Time series of these measurements are publicly available from the SOLIS website at 
http://solis.nso.edu/0/vsm/vsm\_plrfield.html. Future plans include the calculation of the mean polar field
strength from SOLIS/VSM chromospheric observations and the determination of the {\it true} radial polar field 
from SOLIS/VSM full-Stokes measurements.
\end{quote}

\section{Introduction}

Polar field measurements are extremely important for several reasons: 1) they dominate the coronal
structure over much of the solar cycle (except when the polar fields reverse), 2) polar magnetic flux plays a role in 
determining the properties/evolution of the heliospheric magnetic field, 3) the polar magnetic fields are thought 
to be the direct manifestation of the Sun's interior global poloidal fields which serve as seed fields for the global dynamo
that produces the toroidal fields responsible for active regions and sunspots, and 4) the polar regions are the source
of the fast solar wind.

However, measuring the polar field is difficult due to foreshortening effects at the solar
limb as well as the intrinsic weakness of the field near the poles, and interpretation of these measurements
is complicated by a number of factors including the complexity of the
polar magnetic landscape.
Hinode observations of the polar regions have revealed patches of magnetic field with different spatial
extent and distribution. While some are isolated, others form patterns like chains of islands.  Many of these patches
are coherently unipolar and have field strengths reaching above 1 kG.
Their size tends to increase with latitude, up to about $5\times5$ arcseconds.
All of the large patches have fields that are predominantly vertical relative to the local surface, while those of the smaller patches
tend to be horizontal. If a radial correction is applied to line-of-sight (LOS) magnetograms, then
the horizontal fields are incorrectly amplified with a strongly varying radial function. 
Depending on the distribution of the horizontal fields this may lead to a sign bias and inaccurate 
flux on any given day.

Furthermore, for a given latitude, these effects will change with the $B_0$ angle.
Because of projection effects, 
polar measurements obtained at favorable $B_0$ angle (around March/September for the southern/northern solar
hemisphere) will be less noisy than other periods of the year.
The sensitivity of the magnetic field measurement is also a significant factor, and seeing plays a role in ground-based observations.
The impact of all of these factors on time series of polar field measurements is
expected to be greater during solar minimum, when the strength of the poloidal field is stronger.
This document describes the precise procedure used to compute an estimate of the mean polar magnetic field from SOLIS/VSM
measurements.

\section{Wilcox Measurements}

\begin{figure*}[h]
\begin{center}
\includegraphics[width=\linewidth]{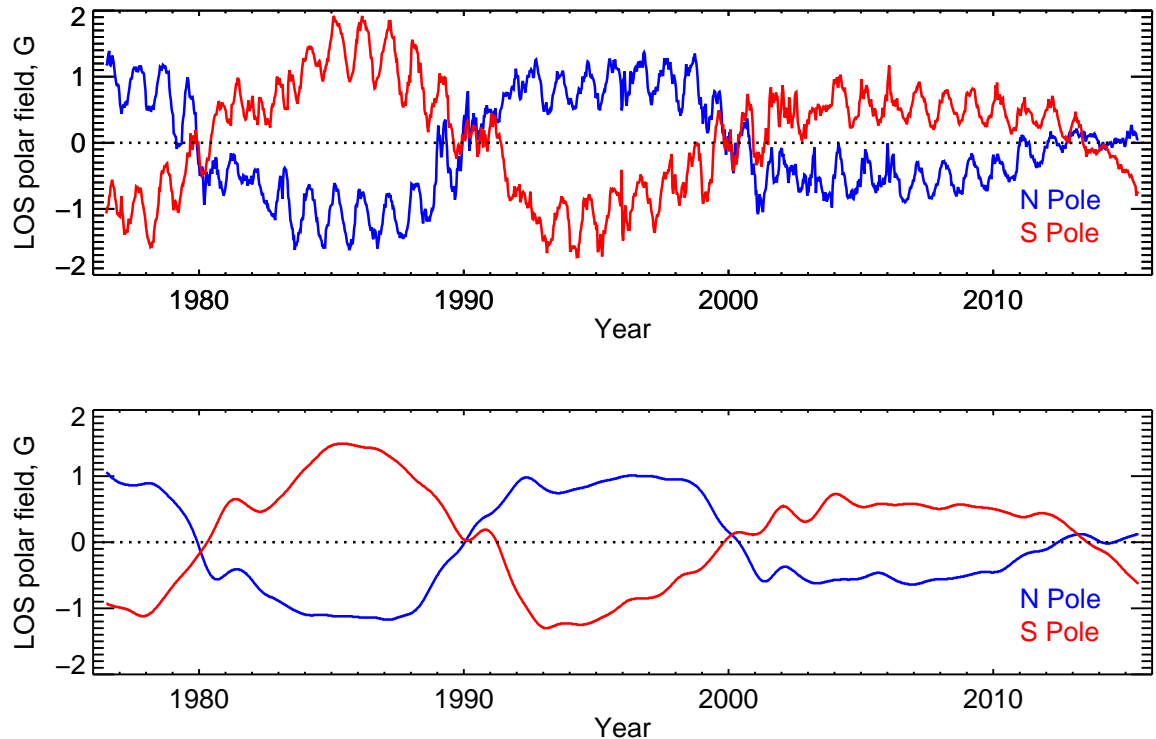}
\caption{Unfiltered (top) and low-pass filtered (bottom) polar field strength measurements taken at Wilcox Solar Observatory
from May 31, 1976 to June 26, 2015.} 
\label{wilc}
\end{center}
\end{figure*}

The Wilcox Solar Observatory (WSO) provides measurements of the polar magnetic fields that data back to May 1976. 
Due to the long history of the program and the homogeneity of the dataset,
they have been used as a reference in many studies.
Daily measurements are made in the north and south hemispheres using 3$^{\prime}$ square apertures that span
the LOS field between approximately $\pm55^{\circ}$ and the corresponding poles.
The solar coordinates of the apertures shift due to
the Earth's orbital motion, and their orientation differs somewhat with each measurement.
Because of the relatively large aperture, the WSO pole measurements are weighted by limb darkening. 
Time series of these measurements are available from the WSO web site at
http://wso.stanford.edu/Polar.html in two flavors: 1) daily values from an average of all
usable measurements in a centered 30-day window and 2) 20nhz low-pass filtered values that eliminate yearly geometric 
projection effects.
Figure \ref{wilc} shows these measurements, in which 
a strong annual modulation due to the variable $B_0$ angle is clearly visible in the unfiltered time series (top).

\section{SOLIS/VSM Photospheric Polar Mean Radial Field}

\subsection{Measurements and Analysis}

The current SOLIS/VSM pipeline provides, among other products, daily full-disk magnetograms of the longitudinal magnetic 
field using the FeI 630.15 nm spectral line.  They are sufficiently high-resolution (1.0 arcseconds beginning in December 2009 and
1.14 prior) to not be significantly weighted by limb darkening.  Before the polar field measurements are made, 
these magnetograms are converted from line-of-sight to radial flux density by assuming that the fields are approximately 
radial at the photosphere. This is a reasonable approximation for network structures and weak fields outside of active regions, 
typical of the solar polar regions of interest.

In general terms, the polar caps typically extend above approximately $60^{\circ}$ latitude in the north and below $-60^{\circ}$
in the south.  For our purposes,
three separate (but overlapping) bands of latitude are considered for each hemisphere: $\pm60^{\circ}$ to $\pm75^{\circ}$,
$\pm60^{\circ}$ to $\pm70^{\circ}$, and $\pm65^{\circ}$ to $\pm75^{\circ}$.
Higher latitudes are not included because they would significantly increase the noise of the 
derived time series.  For all bands, the longitude range is restricted to between $\pm50^{\circ}$.

The mean polar field strength is computed for the selected latitude bands following the approach described
in Bertello et al. 2014 (Solar Physics 289, 2419). 
In short, the original magnetogram pixels are evenly divided into subpixels in order to more accurately
resolve the boundaries of the latitude bands.  Then, the pixels (and partial pixels) corresponding to the latitude band of 
interest are selected according to their computed Stonyhurst heliographic longitudes $(L)$ and latitudes
$(B)$ on the solar disk. If $(x,y)$ are the Cartesian
coordinates of a subpixel relative to the center of the solar disk and the position angle between the geocentric and
solar rotational north poles is zero ($P$=0), then the Stonyhurst heliographic coordinates are given by

\begin{eqnarray}
\sin(B) & = & \sin(B_0)\cos(\rho) + \cos(B_0)\sin(\rho)\cos(\theta) \nonumber \\
\sin(L) & = & -\sin(\rho)\sin(\theta)/\cos(B),
\label{heleqn}
\end{eqnarray}

\noindent where $\rho = \arcsin(r) - S\cdot r$ is the heliocentric angular distance of the subpixel from the
center of the Sun's disk, $r = \sqrt{x^2+y^2}/R_{\circ}$, $\theta$ = arg$(x,y)$, $R_{\circ}$ is the solar
radius in pixels, $S$ is the angular semi-diameter of the Sun, and $B_0$ is the Stonyhurst heliographic latitude
of the observer. 

\begin{figure}[t]
\begin{center}
\includegraphics[width=1.0\textwidth]{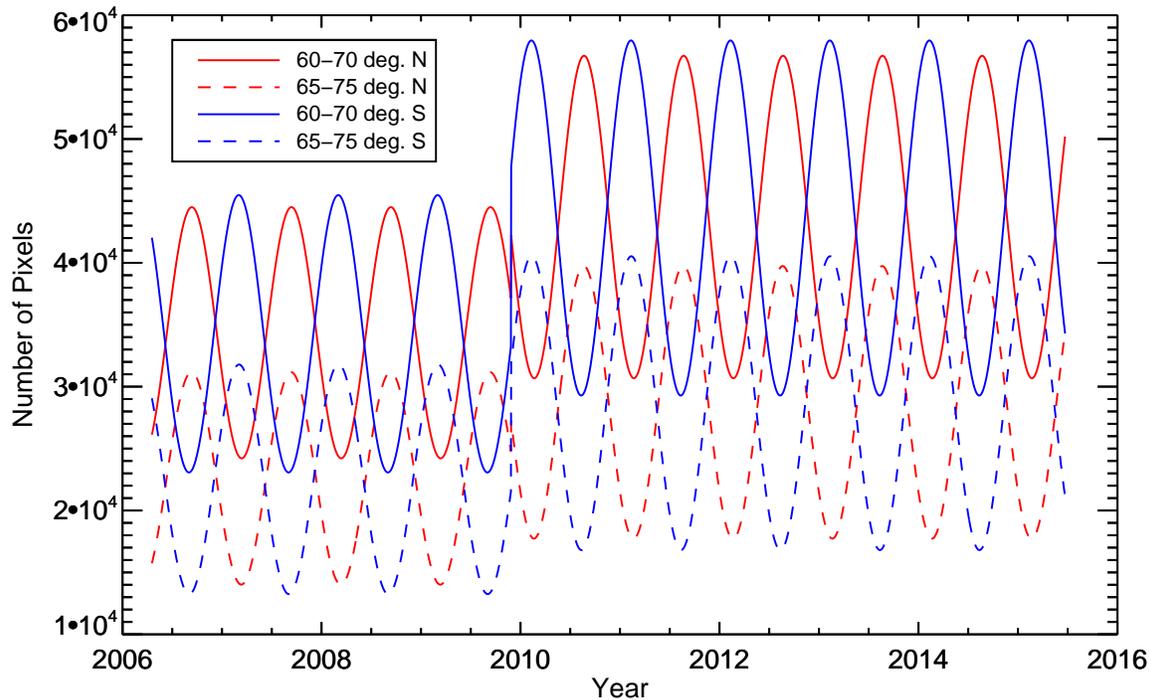}
\caption{Number of full-disk pixels contributing to the calculation of the mean polar field strength
as a function of time for selected latitude bands.}
\label{pixel}
\end{center}
\end{figure}

Figure \ref{pixel} shows how the total number of contributing full-disk pixels (including partial 
pixels) varies as a function
of time for four of the six selected bands. The clearly visible transition in late 2009 corresponds to the
higher spatial resolution of the Sarnoff cameras that replaced the older Rockwell cameras at that time.
The strong annual modulation in the number of pixels is due to the combined
effects of the varying $B_0$ angle and Sun-Earth distance.

If $f_{i}$ is the magnetic radial flux density and $w_i$ is the fractional area ($0 < w \le 1$)
of a pixel $i$ that contributes to a particular latitude band, then the mean polar field strength is
given by

\begin{equation}
B_r = \sum_{i=1}^{N} w_{i} f_{i}~/~\sum_{i=1}^{N} w_{i},
\end{equation}

\noindent where $N$ is the total number of contributors, and the variance is 

\begin{equation}
\sigma^2(B_r) = \frac{1}{N-1}\frac{\sum_{i=1}^{N}w_{i}(f_i - B_r)^2}{\sum_{i=1}^{N}w_{i}}.
\end{equation}

\noindent These quantities were computed for each observation made between May 1st, 2006 and July 14, 2015,
and the corresponding time series for two of the latitude bands 
are shown in Figure \ref{daily} (the other bands exhibit very similar behavior).

\clearpage

\subsection{Results}

\begin{figure}[ht]
\begin{center}
\includegraphics[width=1.0\textwidth]{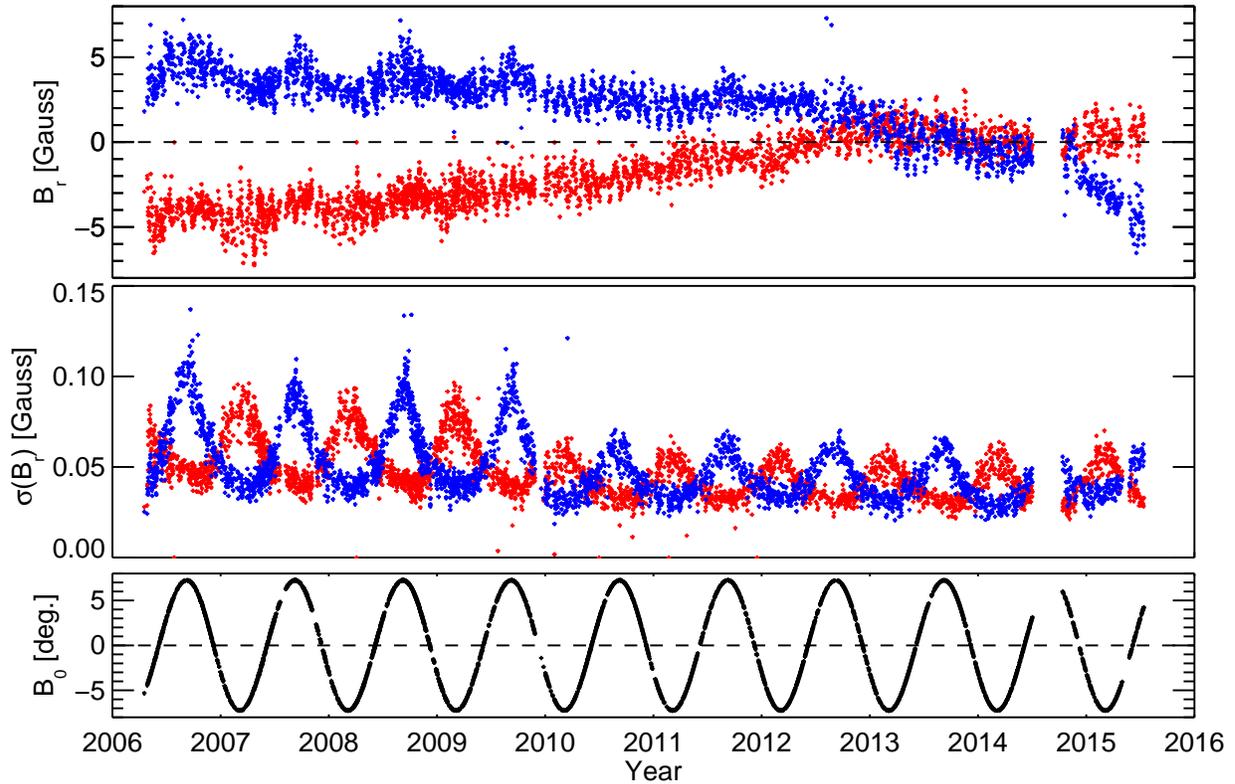}
\caption{Daily SOLIS/VSM unfiltered photospheric mean radial polar field measurements since May 1st, 2006. Top panel shows
north (red) and south (blue) polar measurements computed using heliographic bins in the latitude range [$60^{\circ}$,$75^{\circ}$] 
and [$-60^{\circ}$,$-75^{\circ}$], respectively, and the longitude range [$-50^{\circ}$,$50^{\circ}$].
The corresponding 1-$\sigma$ uncertainty in the daily weighted mean value for each observation is shown
in the middle panel. For reference, the value of the $B_0$ angle for days when observations
were taken is included in the bottom panel.}
\label{daily}
\end{center}
\end{figure}

A quick inspection of Figure \ref{daily} reveals some interesting facts:

\begin{enumerate}
\item The $\sigma$ time series show a strong modulation in phase and anti-phase with the $B_0$ angle. 
For the south measurements, this modulation
is in phase with $B_0$ (\emph{i.e}., the errors are larger when $B_0 > 0$) as expected. 
The opposite is true for the north pole measurements.
Also, measurements taken before December 2009 with the Rockwell cameras show significantly larger errors, 
suggesting that the signal-to-noise ratio is higher in the Sarnoff era.
\item Polar field measurements taken with the older Rockwell cameras show a clear annual modulation. 
This correlates quite well with
the modulation visible in the $\sigma$ time series. 
This could, perhaps, be attributed to the larger uncertainties of these measurements
compared to those obtained after December 2009 with the Sarnoff cameras. Differences in the derived uncertainties 
between the two cameras are expected, due to the different pixel-size and signal-to-noise.
\item A $\sim$27-day periodicity, due to solar rotation, is clearly detected in the VSM polar measurements. 
\end{enumerate}

\begin{figure}[ht]
\begin{center}
\includegraphics[width=1.0\textwidth]{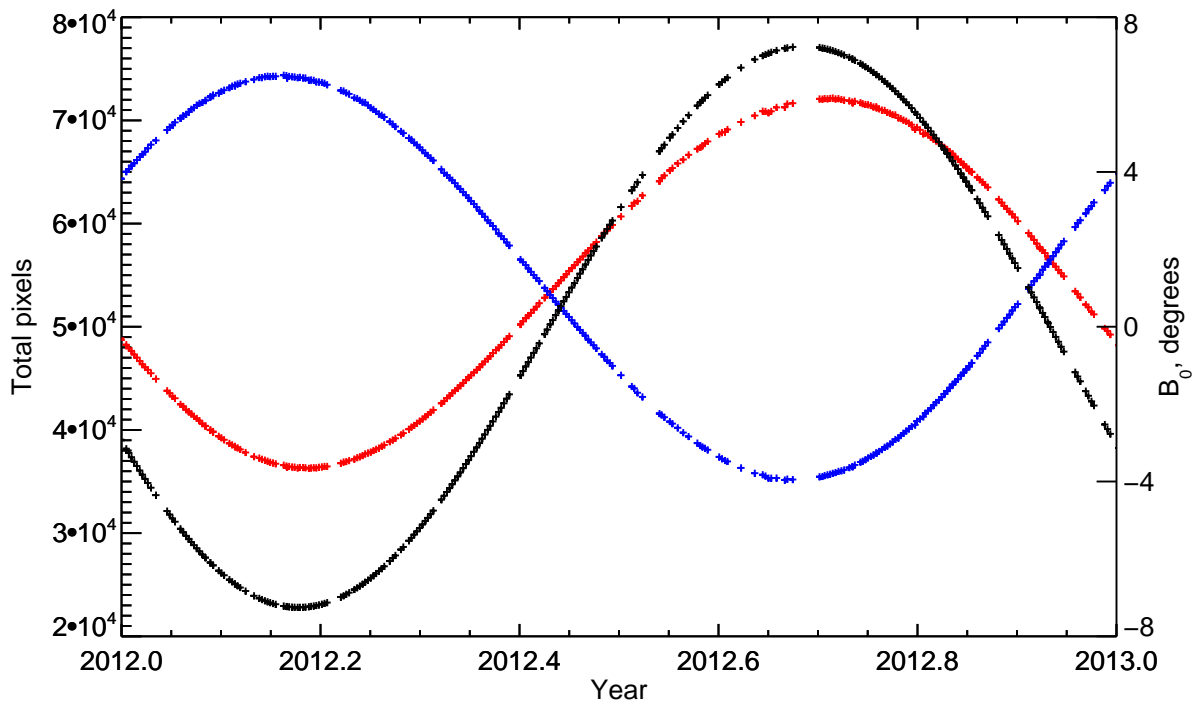}
\caption{Total number of daily contributing full-disk pixels to the [$60^{\circ}$,$75^{\circ}$] north (in red) and
[$-60^{\circ}$,$-75^{\circ}$] south (blue) latitude bands. Also shown are the values
of the $B_0$ angle (black) for days when observations were taken during 2012.}
\label{pixel_2}
\end{center}
\end{figure}

In general, the reliability of magnetic field measurements in the polar regions depends on the value
of the $B_0$ angle. During times of positive $B_0$ angles, the northern polar regions are more easily observable
than the southern regions. The opposite is true when $B_0$ is negative. One can account for this effect by properly weighting
the daily measurements with some quantity that reflects this trend and produce a filtered (smoothed)
version of the mean radial field time series. One such choice for the weights is
the total number of full-disk pixels that contribute to the mean polar field strength of a given latitude band.
Figure \ref{pixel_2} shows the behavior of this quantity
for 2012, both for the north and south hemisphere. The total number of pixels in the north correlates extremely well
with $B_0$, while those in the south are anti-correlated.

\begin{figure}[ht]
\begin{center}
\includegraphics[width=1.0\textwidth]{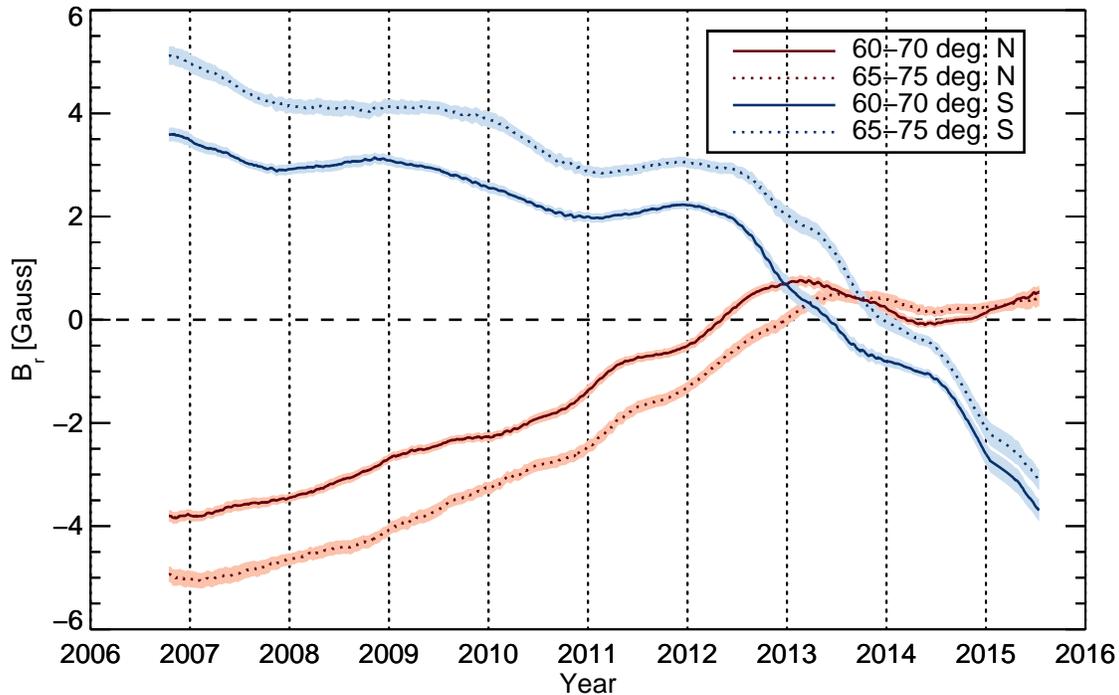}
\caption{Daily SOLIS/VSM filtered photospheric radial polar field measurements from October 15, 2006 to
July 15, 2015. The 3-$\sigma$ error bars are shown as shadowed color areas surrounding the values.
}
\label{filt_180VSM}
\end{center}
\end{figure}

Assuming a 361-day interval centered on a given day $i$, a filtered (smoothed) daily
mean radial polar field value can be calculated as

\begin{equation}
\overline{B}_{r,i} = \sum_{j=i-180}^{i+180}N_jB_{r,j}~/~\sum_{j=i-180}^{i+180}N_j,
\end{equation}

\noindent where $B_{r,j}$ is the daily unfiltered value (from Eq. 2) and $N_j$ is the corresponding total number of full-disk contributing
pixels. The corresponding variance is then

\begin{equation}
\sigma^2(\overline{B}_{r,i}) = \frac{1}{360}\left[\sum_{j=i-180}^{i+180}N_{j}(B_{r,j} - \overline{B}_{r,i})^2~/~\sum_{j=i-180}^{i+180}N_{j}\right].
\end{equation}
Prior to applying the above formulas, multiple measurements taken on the same day were averaged together,
and the daily unfiltered time series were interpolated for missing days. Also, to account for the installation
of new cameras in December 2009, the values of $N_j$ were divided by the square of the camera's 
spatial resolution (\emph{i.e.}, 1 before December 2009 and $1.14^2$ afterwards).
Because the sums in Eqs. 4-5 are necessarily truncated at the edges of the time series, the most recent filtered values are
computed using shorter time intervals. Results for two pairs of latitude bands are shown in Figure \ref{filt_180VSM}. 

\begin{figure}[h]
\begin{center}
\includegraphics[width=1.0\textwidth]{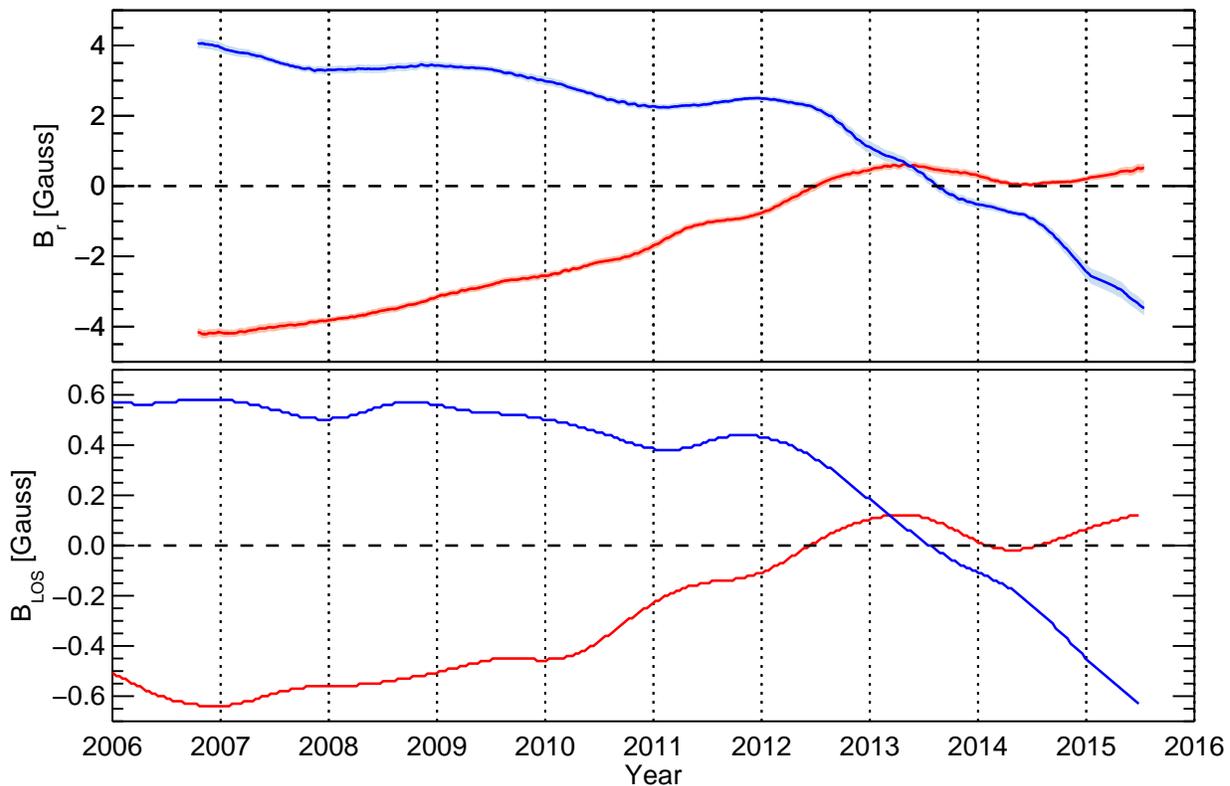}
\caption{Comparison between SOLIS/VSM radial $60^{\circ}-75^{\circ}$ (top) and Wilcox LOS (bottom) filtered polar measurements (note the different
scales). Northern/southern hemisphere
measurements are shown in blue/red, and 3-$\sigma$ error envelopes are included for SOLIS/VSM.}
\label{vsm_wso}
\end{center}
\end{figure}

A close look at this plot shows that during solar cycle 24, the mean polar fields measured in the lower 
latitude bands reverse
their polarity for the first time about eight months before those at higher latitude. While the southern polar caps 
reversed only once, 
the behavior in the northern hemisphere is more complex. After a first reversal, in 2012, the strength of the northern magnetic
polar caps decreased and stay around zero for most of 2014. The northern higher latitude band ($60^{\circ}-70^{\circ}$) never 
quite crossed the zero-line again, but the lower band seems to have changed its polarity at least another
time before its final reversal in late 2015. 
For comparison purposes, we also plot in Figure \ref{vsm_wso} both SOLIS/VSM results for the two broader latitude bands
and the Wilcox filtered LOS polar field measurements. Except for the different scales, the overall
behavior of the VSM and Wilcox time series are very similar. The times of polar reversal in both the northern and southern
hemispheres are also within about a month as indicated in Table 1.

\begin{table}[h]
\begin{center}
\begin{tabular}{|ll|ll|}
\hline
\multicolumn{2}{|c|}{\bf Wilcox} & \multicolumn{2}{c|}{\bf SOLIS/VSM} \\
North & South & North & South \\[0.5em]
08/10/2014 & 07/26/2013 & 07/25/2014 & 08/15/2013  \\[0.5em]
\hline
\end{tabular}
\caption{Approximate times of final polar reversals during solar cycle 24 determined
from the Wilcox and SOLIS/VSM photospheric measurements shown in Figure \ref{vsm_wso}.}
\end{center}
\end{table}

\clearpage

\section{Future Work}

Future plans include calculation of the mean polar field
strength from SOLIS/VSM chromospheric Ca II 854.2 nm observations and determination of the {\it true} radial polar field 
from the SOLIS/VSM photospheric Fe I 630.15 nm full-Stokes data. Although the assumption that the magnetic field
is, on average, radial cannot be made in the chromosphere, determination of the polar field strength from line-of-sight
measurements in the core of the Ca II 854.2 nm spectral line is expected to be more reliable than its photospheric counterpart
due to the canopy effect of magnetic flux tubes in the chromosphere. Estimation of the
polar field strength from full-Stokes data is challenging
due to their limited sensitivity in areas of relatively quiet Sun that are typical in the polar regions. 
However, SOLIS/VSM photospheric vector data at Fe I 630.15 nm uniquely provide both high-spatial and high-spectral
resolution. Because the former is not required for determination of the mean polar field, the
sensitivity of these measurements can be significantly improved by spatially averaging the spectra 
prior to the Milne-Eddington inversion.

Finally, a project is underway to recalibrate older Kitt Peak magnetograms obtained with the 40-channel (Jan. 1970 - March 1974), 512-channel (Jan. 1974 - April 1993), and spectromagnetograph (April 1992 - Sept. 2003) instruments, as well as the earliest
SOLIS/VSM photospheric magnetograms (August 2003 - May 2006). 
Once complete, the polar field strength time series can be extended to a period in excess of 45 years.

\medskip

The authors wish to acknowledge the important contribution to this report by the other members of the SOLIS 
scientific and technical staff who, through useful discussions and suggestions, made this project possible.

\end{document}